\documentclass{ws-procs9x6}
\usepackage{graphicx}

\begin{document}

\title{Hadronic Decays of Baryons in Point-Form Relativistic Quantum Mechanics
}

\author{T. MELDE\footnote{\uppercase{W}ork partially
supported by \uppercase{INFN} and  \uppercase{MIUR-PRIN}.}, 
W. PLESSAS, AND R.~F. WAGENBRUNN}

\address{Theoretische Physik,  Institut f\"ur Physik,
Universit\"at Graz\\
Universit\"atsplatz 5, A-8010 Graz, Austria}

\author{L. CANTON}

\address{INFN, Sezione di Padova, and Dipartimento di Fisica
dell'Universit\`a di Padova\\
Via F. Marzolo 8, I-35131 Padova, Italy}  

\maketitle

\abstracts{
We discuss strong decays of baryon resonances
within the concept of relativistic constituent quark models. In
particular, we follow a  Poincar\'e-invariant approach along the point form
of relativistic quantum mechanics. Here, we focus on pionic decay modes of
$N$ and $\Delta$ resonances. It is found that the covariant quark-model
predictions calculated in the point-form spectator model in general
underestimate the experimental data considerably. This points to a
systematic defect in the used decay operator and/or the baryon wave
functions. From a detailed investigation of the point-form decay operator it
is seen that the requirement of translational invariance implies effective
many-body contributions. 
Furthermore, one has to employ a normalization factor in the definition of
the decay operator in the point-form spectator model. Our analysis suggests
that this normalization factor is best chosen consistently with the one used
for the electromagnetic and axial current operators for elastic 
nucleon form factors.
}
\section{Introduction}
\begin{figure}[b]
{
\includegraphics[width=7.1cm]{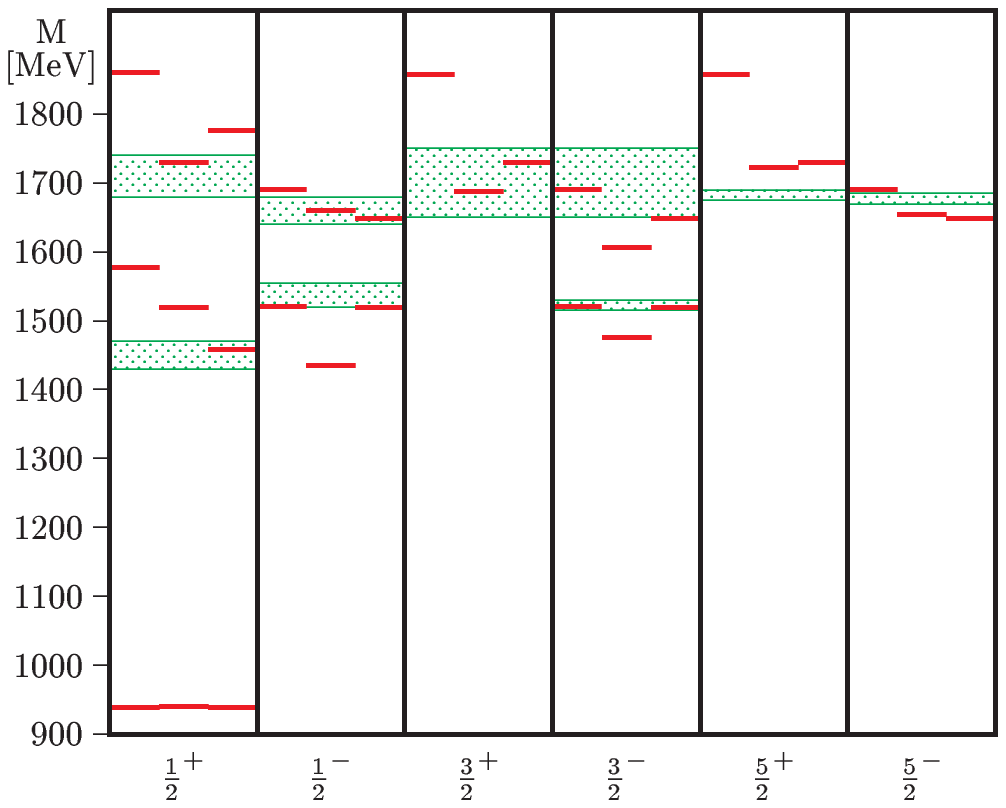}%
\hspace{0.5cm}%
\includegraphics[width=3.76cm]{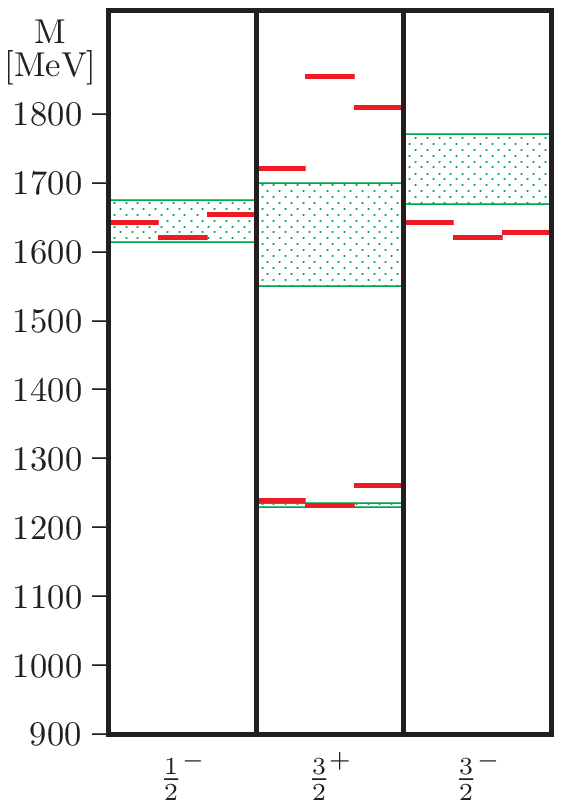}%
}
\caption{Nucleon excitation spectra of three different types of
relativistic CQMs. The left panel shows the nucleon spectrum and
the right panel the $\Delta$ spectrum. 
In each column the left horizontal lines represent the results of the
relativized Bhaduri-Cohler-Nogami CQM\protect\cite{Theussl:2000sj},
the middle ones of the II CQM (Version A)\protect\cite{Loring:2001kx},
and the right ones of the GBE CQM\protect\cite{Glozman:1998ag}.
The shadowed boxes give the experimental
data with their uncertainties after the latest compilation of the
PDG\protect\cite{Eidelman:2004wy}.
}
\label{fig:spectrum}
\end{figure}

Constituent quark models (CQMs) provide an effective description of hadrons
in the low-energy regime of quantum chromodynamics (QCD). It has been of
particular interest to find an appropriate type of (phenomenological)
interaction between the constituent quarks, which is usually comprised of a
confinement and a hyperfine part.
Usually the hyperfine interaction is derived from one-gluon exchange 
(OGE)\cite{Capstick:1986bm}. Beyond that
also alternative types of CQMs have been suggested such as 
the ones based on instanton-induced (II) forces\cite{Loring:2001kx} 
or Goldstone-boson-exchange (GBE) dynamics\cite{Glozman:1998ag}. All these
variants of modern CQMs describe the overall trends of the hadronic spectra 
reasonably well. With regard to baryons, however, only the GBE CQM succeeds
in reproducing simultaneously the level ordering of positive- and 
negative-parity resonances as well as the
$N-\Delta$ splitting in agreement with phenomenology\cite{Glozman:1998fs}.
The typical spectra of the different types of CQMs are exemplified in 
fig.~\ref{fig:spectrum}.

Another problem has been the role of relativity. 
Specifically, it has been found that covariant predictions of 
relativistic CQMs for electroweak nucleon 
form factors agree surprisingly well with experimental data. This is
especially true for the II CQM in a Bethe-Salpeter 
type approach\cite{Merten:2002nz}. The same has been found for the GBE 
CQM and likewise the OGE CQM within the point-form approach of
relativistic quantum
mechanics\cite{Wagenbrunn:2000es,Glozman:2001zc,Boffi:2001zb}. 
In contrast, any nonrelativistic calculation fails drastically.

The covariant results obtained so far for mesonic decay widths of $N$ and
$\Delta$ resonances have shown a systematic underestimation of the
experimental data for any of the three types of 
CQMs\cite{Melde:2002ga,Metsch:2003ix,Melde:2004xj}. Clearly, this points
to shortcomings in the baryon (resonance) wave functions and/or
the decay operators
employed. In this contribution we give a review of the current
status of the investigations on hadronic decays of $N$ and $\Delta$
resonances.

\section{Theory}

Generally, the decay width $\Gamma$ of a resonance is defined by the
expression
\begin{equation}
\label{eq:decwidth}
	\Gamma=2\pi \rho_{f}\left| F\left(i\rightarrow
	f\right)\right|^{2},
\end{equation}
where $ F\left(i\rightarrow f\right)$ is the transition amplitude
and $\rho_{f}$ is the phase-space factor. In eq.~(\ref{eq:decwidth}) 
one has to average over the initial and to sum over the final spin-isospin
projections. A common problem in nonrelativistic approximations of the
transition amplitude is the ambiguity of the proper phase-space 
factor\cite{Geiger:1994kr,Kumano:1988ga,Kokoski:1987is}. 
Here, we present a Poincar\'e-invariant approach, adhering to the 
point form of relativistic quantum mechanics\cite{Keister:1991sb}. 
In this case the generators of the Lorentz transformations remain purely
kinematical and the theory is manifestly covariant\cite{Klink:1998pr}.
This also allows to resolve the ambiguity in
the phase-space factor. The interactions between the constituent quarks
are introduced into the (invariant) mass operator following the
Bakamjian-Thomas construction\cite{Bakamjian:1953}. In this approach
the covariant transition amplitude for mesonic decays is defined via 
the matrix element of the decay operator
\begin{eqnarray}
F\left(i\rightarrow f\right)&=&\left<P',J',\Sigma'\right|{\hat D}_{m}
\left|P,J,\Sigma\right>
\nonumber\\
&=&
\frac{2}{MM'}\sum_{\sigma_i\sigma'_i}\sum_{\mu_i\mu'_i}{
\int{
d^3{\vec k}_2d^3{\vec k}_3d^3{\vec k}'_2d^3{\vec k}'_3
}}
\nonumber\\
&&
\sqrt{\frac{\left(\omega_1+\omega_2+\omega_3\right)^3}
{2\omega_1 2\omega_2 2\omega_3}}
\sqrt{\frac{\left(\omega'_1+\omega'_2+\omega'_3\right)^3}
{2\omega'_1 2\omega'_2 2\omega'_3}}
\nonumber\\
&&
{
\Psi^\star_{M'J'\Sigma'}\left({\vec k}'_i;\mu'_i\right)
\prod_{\sigma'_i}{D_{\sigma'_i\mu'_i}^{\star \frac{1}{2}}
\left\{R_W\left[k'_i;B\left(V'\right)\right]\right\}
}}
\nonumber\\
&&
\left<p'_1,p'_2,p'_3;\sigma'_1,\sigma'_2,\sigma'_3\right|{\hat D}_{m}
\left|p_1,p_2,p_3;\sigma_1,\sigma_2,\sigma_3\right>
\nonumber\\
&&
\prod_{\sigma_i}{D_{\sigma_i\mu_i}^{\frac{1}{2}}
\left\{R_W\left[k_i;B\left(V\right)\right]\right\}
}
\Psi_{MJ\Sigma}\left({\vec k}_i;\mu_i\right)\, ,
\label{transampl}
\end{eqnarray}
where overall momentum conservation,
$P_{\mu}-P'_{\mu}=Q_{\mu}$, is explicitly satisfied; $Q_\mu$ being the 
meson four-momentum.
Here $\left|P,J,\Sigma\right>$ and $\left<P',J',\Sigma'\right|$ are the
eigenstates of the decaying resonance and the outgoing nucleon ground 
state, respectively. The eigenstates are denoted by the eigenvalues 
$P,J,\Sigma$ of the four-momentum operator $\hat P$, the total
angular-momentum operator $\hat J$ and its $z$-component $\hat{\Sigma}$.
The corresponding rest-frame baryon wave functions
$\Psi^{\star}_{M'J'\Sigma'}$
and $\Psi_{MJ\Sigma}$ stem from the velocity-state representations of the
baryon states $\left<P',J',\Sigma'\right|$ and $\left|P,J,\Sigma\right>$. 
The rest-frame quark momenta ${\vec k}_i$, 
for which $\sum_i{{\vec k}_i}=\vec 0$, are related to the individual quark 
four-momenta by the Lorentz boost relations $p_i=B\left(v\right)k_i$, 
with analogous relations holding for the primed variables.

In previous studies of the electroweak nucleon 
structure\cite{Wagenbrunn:2000es,Glozman:2001zc,Boffi:2001zb}
a point-form spectator model (PFSM) for the electromagnetic and axial 
currents performed very well. Consequently, in a first investigation 
of mesonic decays we adopt a PFSM also for the decay operator. 
Assuming a pseudovector quark-meson coupling we express it by
\begin{eqnarray}
& &\left<p'_1,p'_2,p'_3;\sigma'_1,\sigma'_2,\sigma'_3\right|
{\hat D}_{m}\left|p_1,p_2,p_3;\sigma_1,\sigma_2,\sigma_3\right>
=
\nonumber \\
&&
3{N}
\frac{i g_{qq\pi}}{2m_1\left(2\pi\right)^{\frac{3}{2}}}
{\bar u}\left(p'_1,\sigma'_1\right)
\gamma_5\gamma^\mu \lambda_m
u\left(p_1,\sigma_1\right)
Q_\mu
\nonumber \\
&&
2p_{20}\delta\left({\vec p}_2-{\vec p}'_2\right)
2p_{30}\delta\left({\vec p}_3-{\vec p}'_3\right)
 \delta_{\sigma_{2}\sigma'_{2}}
   \delta_{\sigma_{3}\sigma'_{3}}
  \label{eq:decayPFSA}\, ,
  \end{eqnarray}
with the flavor matrix $\lambda_m$ characterizing the particular
decay mode.  
Here, the incoming and outgoing momenta of the active quark are determined
uniquely by the overall momentum conservation of the transition amplitude, 
$P_{\mu}-P'_{\mu}=Q_{\mu}$, together with the two spectator conditions.
Generally, the momentum transferred to the active quark,
$\vec p_1-\vec p_{1}^{\,}{\!}'=\vec {\tilde q}$, is different from the
momentum transfer ${\vec  Q}$ to the baryon as a whole.
It has been shown that this is a consequence of translational invariance
and also induces effective many-body contributions in the definition 
of the spectator-model decay operator\cite{Melde:2004gb}.
Furthermore, in eq.~(\ref{eq:decayPFSA}) there appears an overall
normalization factor $N$. In the PFSM electromagnetic and axial currents
for the nucleon elastic form factors the factor
\begin{equation}
N=\left(\frac{M}{\sum_i{\omega_i}}
\frac{M'}{\sum_i{\omega'_i}}\right)^{\frac{3}{2}}
\label{eq:renorm}
\end{equation}
was employed to recover the proper
proton charge\cite{Wagenbrunn:2000es,Glozman:2001zc,Boffi:2001zb}.
It depends on the individual quark momenta through the $\omega_i$ and
the on-mass-shell conditions of the quarks. 
However, this is not a unique choice, if only Poincar\'e invariance
and charge normalization are imposed. Also any other normalization
factor of the asymmetric form
\begin{equation}
{
N}\left(y\right)=
\left(\frac{M}{\sum_i{\omega_{i}}}\right)^{3y}
\left(\frac{M'}{\sum_i{\omega'_{i}}}\right)^{3\left(1-y\right)}
\label{eq:offsymm}
\end{equation}
would be possible\cite{Melde:2004ce}. Below we also discuss the consequences 
of these further choices of $N$ with regard to the pionic decay widths.
\section{Results}
In table~\ref{tab1} we quote the covariant predictions of the GBE and OGE
CQMs for pionic decay widths calculated with the PFSM decay operator of
eq.~(\ref{eq:decayPFSA}) with the normalization factor of eq.~(\ref{eq:renorm}).
For comparison, we also included the results obtained with the II CQM along
the Bethe-Salpeter approach\cite{Metsch:2004qk}. It is apparent that the
results show a systematic underestimation of the experimental 
data. Only the $N^*_{1535}$ and $N^*_{1710}$ predictions agree 
with the experimental values.
\renewcommand{\arraystretch}{1.5}
%
\begin{table}[b]
\tbl{
PFSM predictions for $\pi$ decay widths of the relativistic
GBE\protect\cite{Glozman:1998ag} and OGE\protect\cite{Theussl:2000sj} CQMs
in comparison to the Bethe-Salpeter results\protect\cite{Metsch:2004qk} 
of the II CQM\protect\cite{Loring:2001kx} and experimental
data\protect\cite{Eidelman:2004wy}. In the last three columns the theoretical
results are expressed as percentage fractions of the (best-estimate)
experimental values in order to be compared to the measured $\Delta \pi$
branching ratios.
\label{tab1}
}
{\begin{tabular}{@{}crccccccc@{}}
\hline
Decays&Experiment&\multicolumn{3}{c}{
Rel. CQM
}&
$\Delta\pi$
&\multicolumn{3}{c}{
$\%$ of Exp. Width
}\\
{$\rightarrow N\pi $}&{[MeV]}& GBE & OGE & II & {\small branching ratio}
& GBE & OGE & II\\
[0.25ex]
\hline
$N^{\star}_{1440}$
&
$\left(227\pm 18\right)_{-59}^{+70}$ &
$33$ & $53$ & $38$ & $20-30\%$ &  $14$ & $24$ & $17$\\ 
$N^{\star}_{1520}$
&
$\left(66\pm 6\right)_{-\phantom{0}5}^{+\phantom{0}9}$&
$17$ & $16$ & $38$ & $15-25\%$ &  $26$ & $24$ & $58$\\ 
$N^{\star}_{1535}$
&
$ \left(67\pm 15\right)_{-17}^{+28}$&
$90$ & $119$ & $33$ & $<1\%$ &  $134$ & $178$ & $49$\\ 
$N^{\star}_{1650}$
&
$\left(109\pm 26 \right)_{-\phantom{0}3}^{+36}$&
$29$ & $41$ & $3$ & $1-7\%$ &  $27$ & $38$ & $3$\\ 
$N^{\star}_{1675}$
&
$ \left(68\pm 8\right)_{-\phantom{0}4}^{+14}$&
$5.4$ & $6.6$ & $4$ & $50-60\%$ &  $8$ & $10$ & $6$\\ 
$N^{\star}_{1700}$
&
$ \left(10\pm 5\right)_{-\phantom{0}3}^{+\phantom{0}3}$&
$0.8$ & $1.2$ & $0.1$ & $>50\%$ &  $8$ & $12$ & $1$\\ 
$N^{\star}_{1710}$
&
$\left(15\pm 5\right)_{-\phantom{0}5}^{+30}$&
$5.5$ & $7.7$ & $n/a$ & $15-40\%$ &  $37$ & $51$ & $n/a$\\ 
$\Delta_{1232}$
&
$\left(119\pm 1 \right)_{-\phantom{0}5}^{+\phantom{0}5}$&
$37$ & $32$ & $62$ & $-$ &  $31$ & $27$ & $52$\\ 
$\Delta_{1600}$
&
$\left(61\pm 26\right)_{-10}^{+26}$&
$0.07$ & $1.8$ & $n/a$ & $40-70\%$ &  $\approx 0$ & $3$ &
$n/a$\\ 
$\Delta_{1620}$
&
$\left(38\pm 8\right)_{-\phantom{0}6}^{+\phantom{0}8}$&
$11$ & $15$ & $4$ & $30-60\%$ &  $29$ & $39$ & $11$\\ 
$\Delta_{1700}$
&
$\left(45\pm 15\right)_{-10}^{+20}$&
$2.3$ & $2.3$ &  $2$ & $30-60\%$ &  $5$ & $5$ & $4$\\
\hline
\end{tabular}}
\end{table}

We also present the numerical results for the decay widths with specific
attention to the $\Delta\pi$ branching ratios, for the magnitudes of
which a striking relationship is found to the underestimation of the
experimental value: the larger the $\Delta\pi$ branching ratio of a
resonance, the smaller the relative prediction for the $\pi$ decay width.
At this point it is not yet clear, if this behaviour is purely accidental
or substantiates a shortcoming of the present description of baryon 
decays. In particular, additional degrees of freedom might be
missing.

\begin{figure}[t]
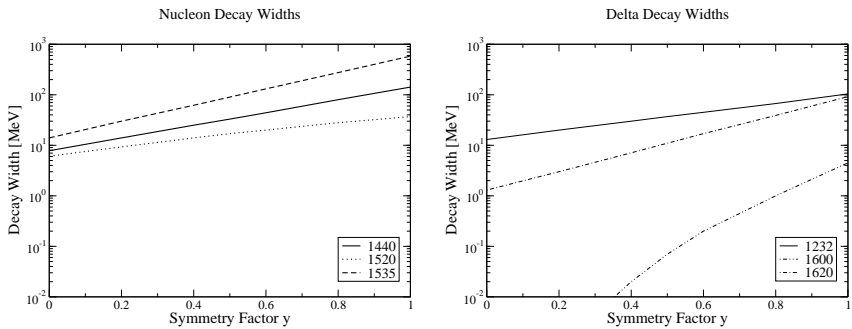

{
\includegraphics[width=5.4cm]{nucleon_off.eps}
\hspace{0.3cm}%
\includegraphics[width=5.4cm]{delta_off.eps}
}
\caption{Dependence of the $\pi$ decay widths on the asymmetry parameter
$y$ in the normalization factor of eq. (\ref{eq:offsymm}) for selected
$N$ and $\Delta$ resonances.
}\label{fig:symmfac}
\end{figure}

Let us now examine the influence of the normalization factor $N$ in
eq.~(\ref{eq:decayPFSA}). In fig.~\ref{fig:symmfac} we show the results
for different possible choices of $N(y)$ according to
eq.~(\ref{eq:offsymm}) in case of the GBE CQM. It is seen that in the
range $0\le y\le 1$ all decay widths grow rapidly with increasing $y$.
Thus one could enhance or reduce the results as compared to the ones
obtained with the symmetric factor $N(y=\frac{1}{2})$. While all other 
choices of $N(y)$ are also allowed, the symmetric factor was used for the 
electroweak nucleon form factors in
refs.\cite{Wagenbrunn:2000es,Glozman:2001zc,Boffi:2001zb}
and likewise the results given in table \ref{tab1}.
We consider this as the best choice also for the baryon decays.
It has the property that no theoretical result overshoots the
corresponding experimental value. This is welcome and can be 
considered as reasonable, since one may expect further contributions
not yet included in the present decay operator to raise 
the theoretical predictions.

\begin{figure}[t]
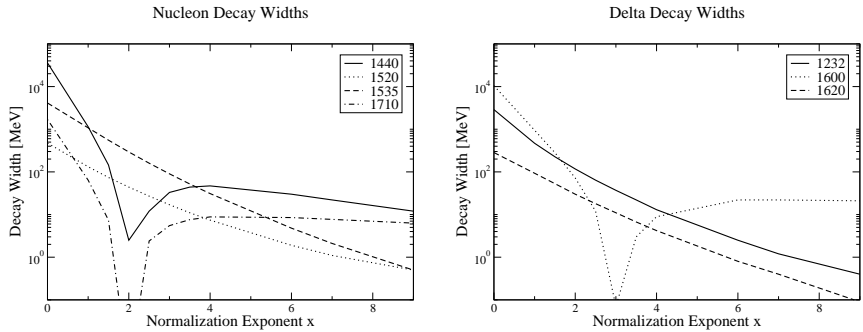

\includegraphics[width=5.4cm]{nucleon1.eps}
\hspace{0.3cm}
\includegraphics[width=5.4cm]{delta1.eps}
\caption{Dependence of the $\pi$ decay widths on the exponent $x$ 
of the normalization factor in eq. (\ref{eq:exponent}) for selected
$N$ and $\Delta$ resonances. 
}\label{fig:exponent}
\end{figure}

The normalization factor $N$ in eq.~(\ref{eq:decayPFSA})
effectively introduces a momentum cut-off. This can even better be
seen if we investigate the form
\begin{equation}
\label{eq:exponent}
{N}\left(x\right)
=
\left(\frac{M}{\sum_i{\omega_{i}}}\frac{M'}{\sum_i{\omega'_{i}}}\,
\right)^{\frac{x}{2}}\, 
\end{equation}
with an arbitrary exponent $x$. It still represents a 
Poincar\'e-invariant construction but it does not guarantee for the 
proper charge normalization unless $x=3$. It is clearly seen in 
fig.~\ref{fig:exponent} that the choice
$x=0$ representing the bare case without normalization factor
yields unreasonable results. The theoretical decay widths then evolve 
smoothly with increasing exponent $x$.  For certain resonances 
(namely,  $N^\star_{1440}$, $N^\star_{1710}$, and $\Delta_{1600}$)
the decay widths have a minimum.  We notice that these are known
as so-called structure-dependent resonances\cite{Koniuk:1979vy}.
They are the radial excitations of the $N$ and $\Delta$ ground states,
respectively, with a corresponding nodal 
behaviour in their wave functions. These characteristics are quite 
distinct from the other resonances, which show a monotonous 
dependence on the exponent $x$. Through this study we get interesting
insights into the causes for occurrence of narrow decay widths,
at least for baryon states with nodal behaviour.

If we assume again the criterion that the theoretical predictions for 
decay widths with the decay operator~(\ref{eq:decayPFSA}) should not 
exceed the experimental data, we find that the case with $x=3$
(and $y=\frac{1}{2}$) is the
optimal choice. This leads to a single postulate for normalizing
the PFSM operators for strong and electroweak processes.
\section{Summary}
We have discussed a Poincar\'e-invariant description of strong baryon 
resonance decays in point-form relativistic quantum mechanics.
Covariant predictions of relativistic CQMs have been
shown for $\pi$ decay widths. They are considerably different from 
previous nonrelativistic results or results with relativistic 
corrections included. The covariant results calculated with a 
spectator-model decay operator show a uniform trend.
In almost all cases the corresponding theoretical predictions
underestimate the experimental data considerably. This is true in the
framework of Poincar\'e-invariant quantum mechanics (here in point form)
as well as in the Bethe-Salpeter approach\cite{Metsch:2004qk}.
Indications have been given that for a particular resonance the size
of the underestimation is related to the magnitude of the $\Delta\pi$ 
branching ratio. This hints to systematic shortcomings in the description 
of the decay widths. 

The investigation of different possible choices for the normalization 
factor occurring in the spectator-model decay operator has led to the 
suggestion that the symmetric choice is the most natural one. It is 
also consistent with the same (symmetric) choice adopted
before for the spectator-model currents in the study of the electroweak
nucleon form factors.

\vspace{-5mm}
\section*{Acknowledgements}
This work was supported by the Austrian Science Fund (Project 
P16945). T.~M. would like to thank the 
INFN and the Physics Department of the University of Padova for their 
hospitality and MIUR-PRIN for financial support.

%

\vspace{-3mm}
\begin{thebibliography}{10}

\bibitem{Capstick:1986bm}
S. Capstick and N. Isgur, Phys. Rev. D {\bf 34},  2809  (1986).

\bibitem{Loring:2001kx}
U. Loering, B.~C. Metsch, and H.~R. Petry, Eur. Phys. J. {\bf A10}, 395
(2001); ibid. 447 (2001).

\bibitem{Glozman:1998ag}
L.~Y. Glozman, W. Plessas, K. Varga, and R.~F. Wagenbrunn, Phys. Rev. D
{\bf 58},  094030  (1998).

\bibitem{Glozman:1998fs}
L.~Y. Glozman {\it et~al.}, Phys. Rev. C {\bf 57},  3406  (1998).

\bibitem{Theussl:2000sj}
L. Theussl, R.~F. Wagenbrunn, B. Desplanques, and W. Plessas, Eur. Phys. J.
{\bf A12}, 91 (2001).
  
\bibitem{Eidelman:2004wy}
S. Eidelman {\it et~al.}, Phys. Lett. {\bf B592}, 1 (2004).

\bibitem{Merten:2002nz}
D. Merten {\it et~al.}, Eur. Phys. J. {\bf A14},  477  (2002).

\bibitem{Wagenbrunn:2000es}
R.~F. Wagenbrunn {\it et~al.}, Phys. Lett. {\bf B511},  33  (2001).

\bibitem{Glozman:2001zc}
L.~Y. Glozman {\it et~al.}, Phys. Lett. {\bf B516},  183  (2001).

\bibitem{Boffi:2001zb}
S. Boffi {\it et~al.}, Eur. Phys. J. {\bf A14},  17  (2002).

\bibitem{Melde:2002ga}
T. Melde, W. Plessas, and R.~F. Wagenbrunn, Few-Body Syst. Suppl. {\bf 14},
37 (2003).

\bibitem{Metsch:2003ix}
B. Metsch, U. Loring, D. Merten, and H. Petry, Eur. Phys. J. {\bf A18},
189 (2003).

\bibitem{Melde:2004xj}
T. Melde, W. Plessas, and R.~F. Wagenbrunn, Contribution to the N*2004
Workshop, Grenoble, to appear in the Proceedings; hep-ph/0406023 (2004).

\bibitem{Geiger:1994kr}
P. Geiger and E.~S. Swanson, Phys. Rev. D {\bf 50},  6855  (1994).

\bibitem{Kumano:1988ga}
S. Kumano and V.~R. Pandharipande, Phys. Rev. D {\bf 38},  146  (1988).

\bibitem{Kokoski:1987is}
R. Kokoski and N. Isgur, Phys. Rev. D {\bf 35},  907  (1987).

\bibitem{Keister:1991sb}
B.~D. Keister and W.~N. Polyzou, Adv. Nucl. Phys. {\bf 20},  225  (1991).

\bibitem{Klink:1998pr}
W.~H. Klink, Phys. Rev. C {\bf 58},  3587  (1998).

\bibitem{Bakamjian:1953}
B. Bakamjian and L.~H. Thomas, Phys. Rev. {\bf 92},  1300  (1953).

\bibitem{Melde:2004gb}
T. Melde, L. Canton, W. Plessas, and R.~F. Wagenbrunn, hep-ph/0411322
(2004).

\bibitem{Melde:2004ce}
T. Melde, L. Canton, W. Plessas, and R.~F. Wagenbrunn, Contribution to
the Mini-Workshop on Quark Dynamics, Bled, to appear in the Proceedings;
hep-ph/0410274 (2004).

\bibitem{Metsch:2004qk}
B. Metsch, Contribution to the 10th International Conference on Hadron
Spectroscopy, Aschaffenburg, 2003; hep-ph/0403118 (2004).

\bibitem{Koniuk:1979vy}
R. Koniuk and N. Isgur, Phys. Rev. D {\bf 21}, 1868 (1980).

\end{thebibliography}

%
\end{document}